%% file: g1_aaa.tex
\documentclass{aa}

\newcommand\aastex{AAS\TeX}

\newcommand\ms{~m~s$^{-1}$}
\newcommand\kms{~km~s$^{-1}$}

\newcommand{\B}[1]{#1}
\newcommand{\PG}{P/2016~G1}
\usepackage{xcolor}
\usepackage{url}
\bibpunct{(}{)}{;}{a}{}{,} % to follow the A&A style
\begin{document}

\title{Disintegration of Active Asteroid P/2016 G1 (PANSTARRS)}

%\watermark{DRAFT}
 
\author{
Olivier R. Hainaut\inst{\ref{inst:eso}}\and
Jan T. Kleyna\inst{\ref{inst:ifa}}\and 
Karen J. Meech\inst{\ref{inst:ifa}}\and 
Mark Boslough\inst{\ref{inst:mb}}\and 
Marco Micheli\inst{\ref{inst:esa},\ref{inst:roma}}\and 
Richard Wainscoat\inst{\ref{inst:ifa}}\and 
Marielle Dela Cruz\inst{\ref{inst:ifa}}\and 
Jacqueline V. Keane\inst{\ref{inst:ifa}}\and 
Devendra K. Sahu\inst{\ref{inst:india}}\and 
Bhuwan C. Bhatt\inst{\ref{inst:india}}
}

\institute{
{European Southern Observatory,
Karl-Schwarzschild-Strasse 2,
D-85748 Garching bei M\"unchen, Germany}\label{inst:eso}
\and
{Institute for Astronomy,
2680 Woodlawn Drive,
Honolulu, HI 96822, USA}\label{inst:ifa}
\and
{University of NM - 1700 Lomas Blvd, NE. Suite 2200, Albuquerque, NM 87131-0001 USA}\label{inst:mb}
\and
{ESA SSA-NEO Coordination Centre,
Largo Galileo Galilei, 
1 00044 Frascati (RM), Italy}\label{inst:esa}
\and
{INAF - Osservatorio Astronomico di Roma,
Via Frascati, 33,
00040 Monte Porzio Catone (RM), Italy}\label{inst:roma}
\and
{Indian Inst. Astrophys., II Block, 
Koramangala, Bangalor 560 034, India}\label{inst:india}
}

\date{Received 10/05/2019 /Accepted 01/07/2019}

%=============================================================================================
\abstract{
We report on the catastrophic disintegration of \PG~(PANSTARRS), an active asteroid, in April 2016. Deep images over three months show that object is constituted by a central concentration of fragments surrounded by an elongated coma, and presents previously unreported sharp arc-like and a narrow linear features. The morphology and evolution of these characteristics independently point toward a brief event on 2016 March 6. The arc and the linear feature can be reproduced by large particles on a ring, moving at $\sim 2.5$\ms\B{. The expansion of the ring defines} a cone with a $\sim 40^\circ$ half-opening. We propose that the \PG\ was hit by a small object which caused its (partial or total) disruption, and that the ring corresponds to large fragments ejected during the final stages of the crater formation.}

          
\titlerunning{\aastex\ P/2016 G1 PANSTARRS}
\authorrunning{Hainaut et al.}

\maketitle

%=============================================================================================

\section{Introduction}
\label{sec:intro}

\PG~was discovered by the Pan-STARRS1 (PS1) telescope on 2016 Apr.~1, and the orbit ($a$ = 2.582 au, $e$ = 0.210, $i$ = 10.969$^{\circ}$, $T_{\rm Jup}$ = 3.367), suggested that this was possibly a main belt comet \citep{weryk2016}.  The object was approaching a perihelion of $q$ = 2.041 au on 2017 Jan. 26.24. Because the surrounding dust coma had a peculiar appearance in the PS1 images, we undertook an immediate campaign to follow this up with images from the 3.6m Canada-France-Hawaii Telescope (CFHT). The first CFHT images, obtained on 2016 Apr. 3 showed a structure to the south, 90 degrees from the dust tail.  Because of the unusual appearance, we dedicated time from our CFHT PS1 discovery follow up program to image \PG\ frequently during the dark runs when the wide field imager was on the telescope.  By mid-April, its diffuse and core-less appearance made it clear that the object was undergoing some sort of catastrophic disruption. 

\citet{hsieh18} reported that \PG\ is linked to the Adeona family, which likely originated in a cratering event $\sim 700$~Myr ago \citep{benavidez12, carruba16, milani17}. The Adeona family members are typically C- and Ch-type, while the surrounding background objects are predominantly of the S-type \citep{hsieh18}.

\section{Observations \& Data Reduction}
\label{sed:observations}

\subsection{Pan-STARRS1}
\label{sec:PS1}

With a well-known orbit and a brightness estimated to be above the PS1 limiting magnitude over most of its orbit, we conducted a search through the PS1 database for pre-discovery images. On 2016 Mar. 7 the object is clearly visible, apparently mostly stellar (see below for a more detailed assessment).  
{Photometry was obtained using PS1's intrinsic calibration.}
%Photometry was obtained on these images using $R$-band standards \R{The image was acquired in w1 - we need to discuss the effect of the R calibration - ideally report everything in R or r but not a mix}, and is shown in Table~\ref{tab-data}.  
A set of four $w$-band (\B{central wavelength} $\lambda = 6250$~\AA, width 4416~\AA) images on 2016 Jan. 10 shows nothing at all, with a limiting magnitude in the stack of $r\sim$23. 
This corresponds to a nucleus radius of 0.81 km for an albedo of 4\% or 0.32 km for an albedo of 25\%.

The position of \PG\ was imaged 12 additional times in the PS1 data prior to discovery, between 2011 Feb. 23 and 2015 Jan. 17 and nothing was visible down to the limiting magnitudes shown in Table~\ref{tab-data}. The most constraining observations (2012 Jun.) suggest that the object must have a radius $R_N<0.4$~km for a 4\% albedo or $R_N<0.2$~km for 25\% albedo. For these non-detections, a positional uncertainty region of 3-$\sigma$ was inspected.  These limits are thus valid under the assumption that the existing astrometric arc is good and that a gravity-only solution models the orbit well enough over 5 years.

%-V-FIG-V----------------------------------------------
\begin{figure}[ht]
\includegraphics[width=8cm]{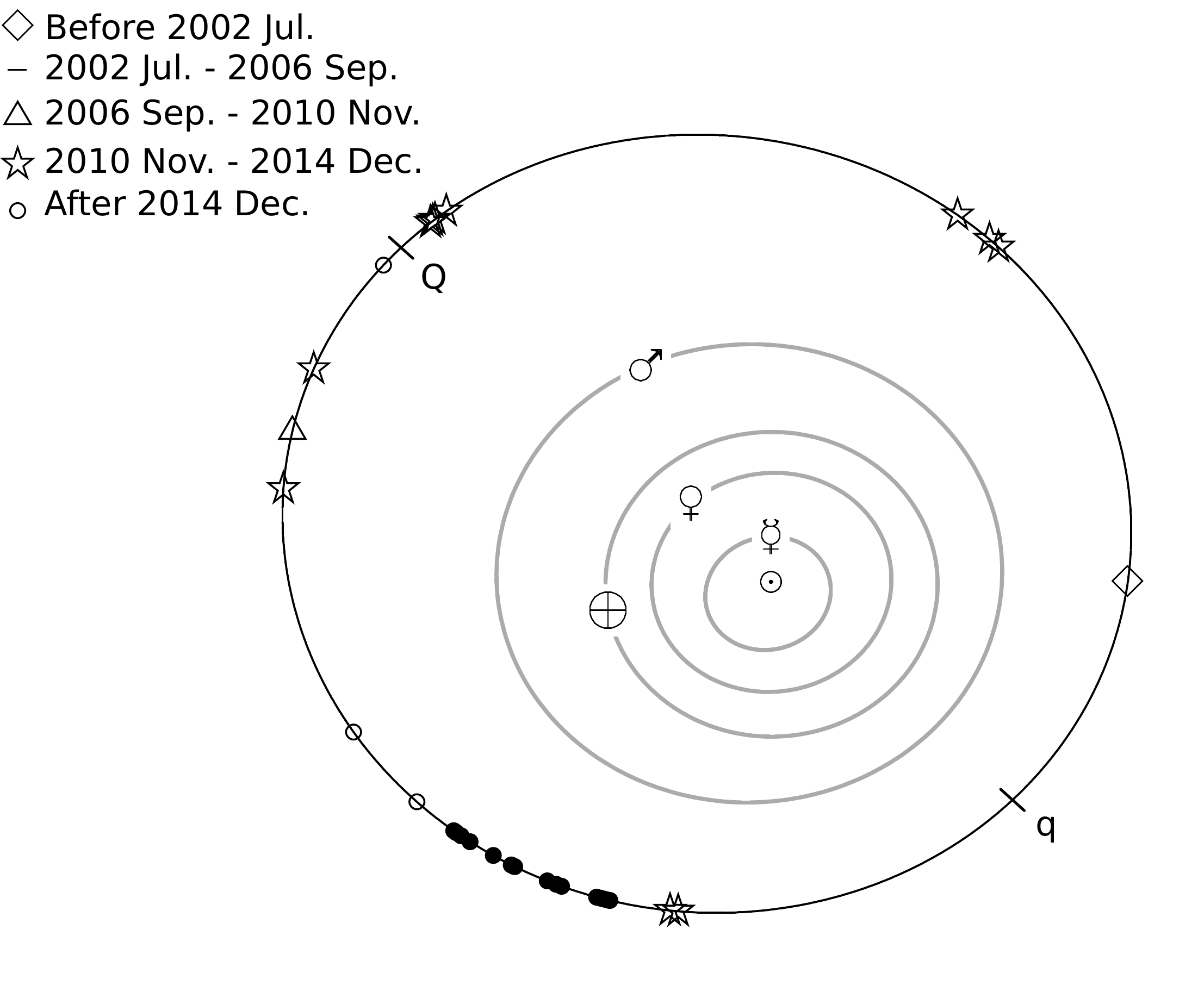}
\caption{
Distribution of observations (see Table 1) of \PG\ along its orbit. The symbols indicate the apparitions, from aphelion (Q) to aphelion. Perihelion is marked (q). Filled symbols show when the comet was detected, open symbols where it was not detected. \B{No observations were acquired during the 2002 Jul. -- 2006 Sep. apparition}. The positions of the planet symbols correspond to 2016 Jul.
%Fig: Jan, edited in inkscape
}
\label{fig:orbit}
\end{figure}
%-----------------------------------------------

%--FIG-2--v----------------------------------------------
\begin{figure*}[ht!]
\begin{center}
\includegraphics[width=17.5cm]{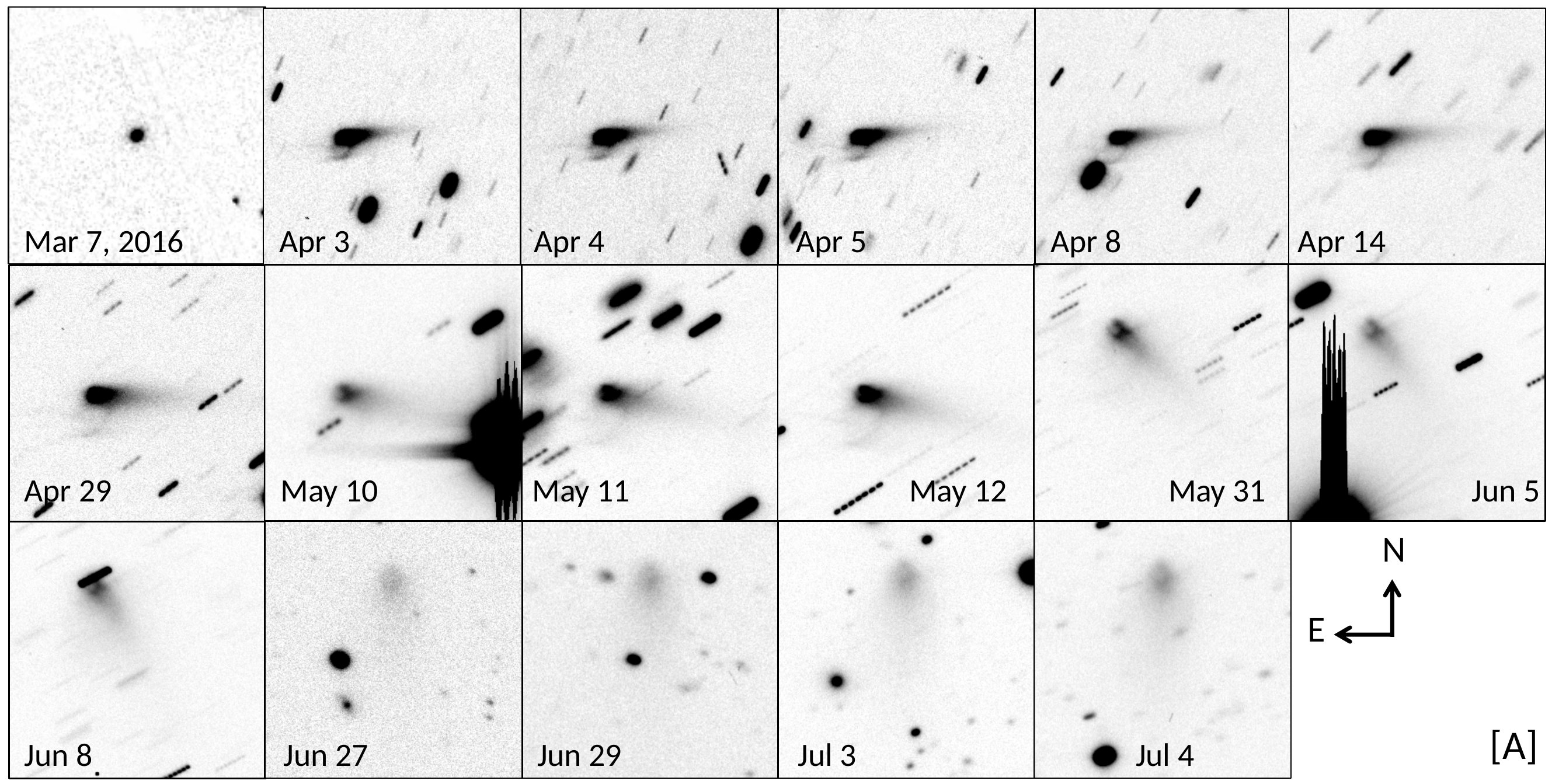}
\includegraphics[width=17.5cm]{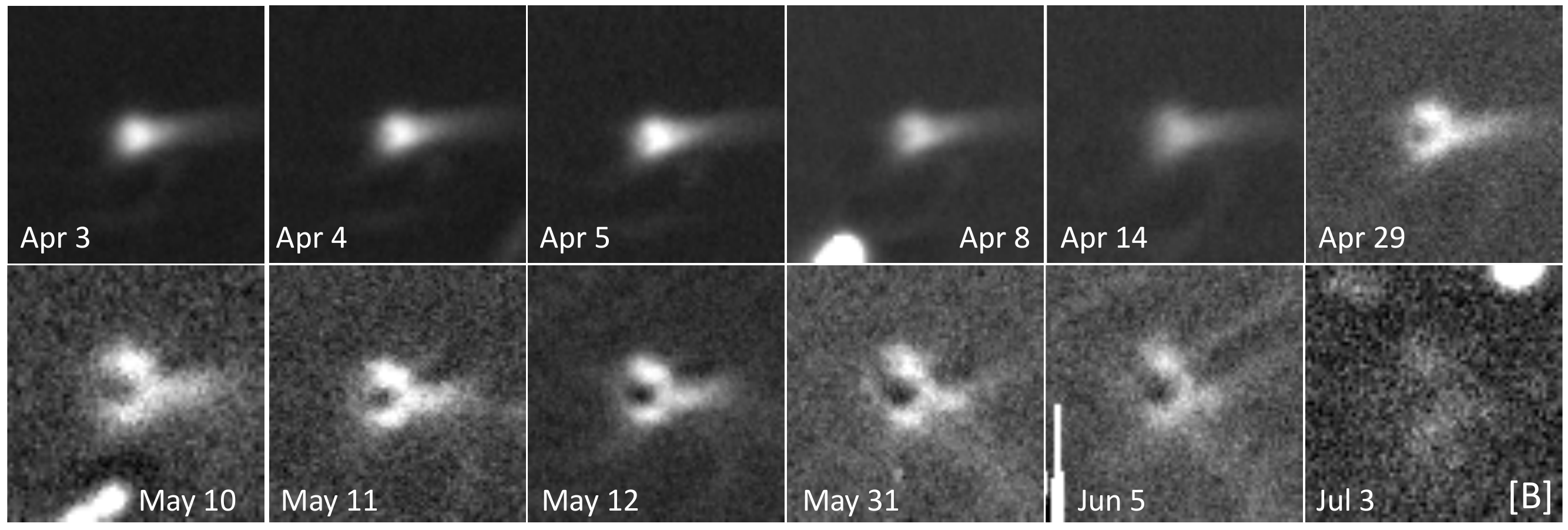}
\includegraphics[width=17.5cm]{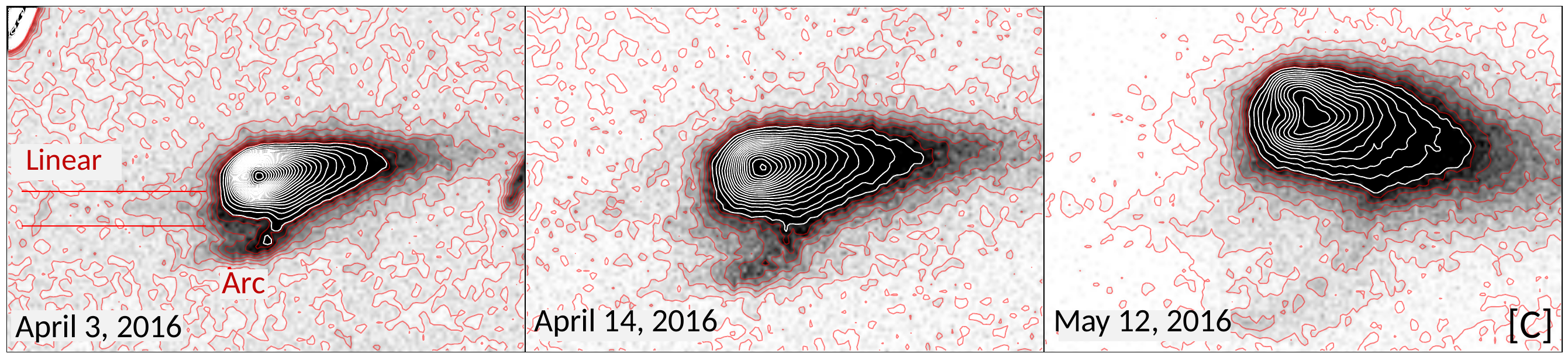}
\caption{\small
[A] Images of P/2016 G1 between UT 2016 March 7 and July 4.  All the images are scaled to be 6$\times$10$^4$ km on a side. The Earth crossed the object's orbital plane on Apr. 13, so images around that date show the coma extension above and below the plane. 
[B] Images of P/2016 G1 that have been processed with unsharp masking to enhance 
\B{the three clumps and the details} near the disintegrating body.  Each image is 1.5$\times$10$^4$ km on a side.
[C] Contour plots on three dates showing the expansion of the southern arc and the eastern linear feature (highlighted with red lines). Each panel is $30''\times 20''$. The panel widths and pixel scales are: April 3 (34,000~km, 227~km), April 14 (32,000~km, 231~km) and May 12 (29,000~km, 193~km). North is up, East is Left. The Sun and velocity vectors are listed in Table~1.
}
\label{fig:images}
\end{center}
\end{figure*}
%-----^------------------------------------------

\subsection{Canada-France-Hawaii Telescope}
Imaging data were obtained using the MegaCam wide-field imager which is an array of 40 2048$\times$4612 pixel CCDs, with a plate scale of 0$\farcs$187 pixel$^{-1}$.  The epochs and circumstances of the CFHT data are listed in Table~\ref{tab-data}. MegaCam is on the telescope for a period centred on each new moon, and data are obtained through queue service observing and are processed to remove the instrumental signature through the Elixir pipeline\footnote{http://www.cfht.hawaii.edu/Instruments/Elixir}.

The photometric calibration of the processed data accesses the Pan-STARRS database or SDSS to provide a photometric zero point to each frame, using published colour corrections \citep{tonry12} {to translate PS1 $g,r,i,z$ bands into SDSS or Johnson-Cousins} systems.
The object headers are used to identify the target and download orbital elements from the Minor Planet Center; the resulting object location is used to determine which object in the frame corresponds to the target.  In the final pass, Terapix tools \citep[SExtractor,][]{SEx} are run to produce multi-aperture and automatic aperture target photometry.

A first series of CFHT observations was obtained as an immediate follow-up to the discovery, showing the dust features that constitute the core of the following discussion. A second was acquired with the broad $gri$ filter on 2018 Dec. 12 and 31, over 1000~days after the first series. The object has completely dispersed, showing no visible remain down to $gri=26$ ($5\sigma$ point-source, using the MegaCam Direct Imaging Exposure Time Calculator\footnote{http://cfht.hawaii.edu/Instruments/Imaging/  MegaPrime/dietmegacam.html}. 

\subsection{Himalayan Chandra Telescope (HCT)}

Observations of P/2016 G1 were made on 2016 May 31 using the Himalayan Chandra 2.0-m telescope with the Himalaya Faint Object Spectrograph and Camera (HFOSC). The instrument uses a SITe ST-002 CCD with a plate scale of 0$\farcs$296 per pixel.  Data were obtained through Bessell $R$-band filters under clear conditions with seeing of $\sim$1$\farcs$6.

We processed the data using our image reduction pipeline which bias-subtracts and flattens the data, and then applies the Terapix tools to fit a precise WCS to the frame. Following the removal of instrumental effects, the photometric calibration proceeded in the same manner as for the CFHT data.

\subsection{Observatory Archives and other publications}
\label{sec:archives}

Additional pre-discovery images were found from the CFHT (2007 Feb. 16) and the INT on La Palma (2000 Oct. 20), both with non-detections, both less constraining than the 2012 Jun. images. For these older images, the positional uncertainties were $\pm$45$''$ and $\pm$3$'$, respectively.  

Finally, the Canadian Astronomy Data Centre's Solar System Object Image Search \citep{soss} was used, and about 40 short exposures with the Cerro Tololo Interamerican Observatory 4-meter telescope were identified. Based on the 2016 Jan. 10 upper limit, and accounting for the geometry of the orbit, these CTIO images would fail to reach the same limit by 1--2~mag.

The positions on the orbit for the dates for which we have observations (including both detections and non-detections) are shown in Fig.~\ref{fig:orbit}. This shows that we have tightly bracketed the period of apparent activity for this current apparition, strongly suggesting that whatever caused the activity began between 2016 Jan. 10 and 2016 Apr. 1 when P/2016 G1 was discovered, and that the object did not present significant activity during previous apparitions.

\cite{moreno2016} have acquired images with the 10.4~m Gran Telescopio Canarias on 2016 Apr. 21, May 29 and Jun. 08. The general appearance of their images fits well with our CFHT images discussed below. They followed up with two observations obtained with the Hubble Space Telescope on 2016 Jun. 28 and Jul. 11 \citep{moreno2017}, which focus on the central region of the disrupted object. From these deep, high-resolution images, they estimate that no fragment larger than $\sim 30$~m survived. They modelled the observations using their Monte Carlo dust-tail code; their results are discussed in the comparison with ours in Section~\ref{sec:moreno}.

%-V-FIG-V----------------------------------------------
\begin{figure}[ht!]
\includegraphics[width=8.5cm]{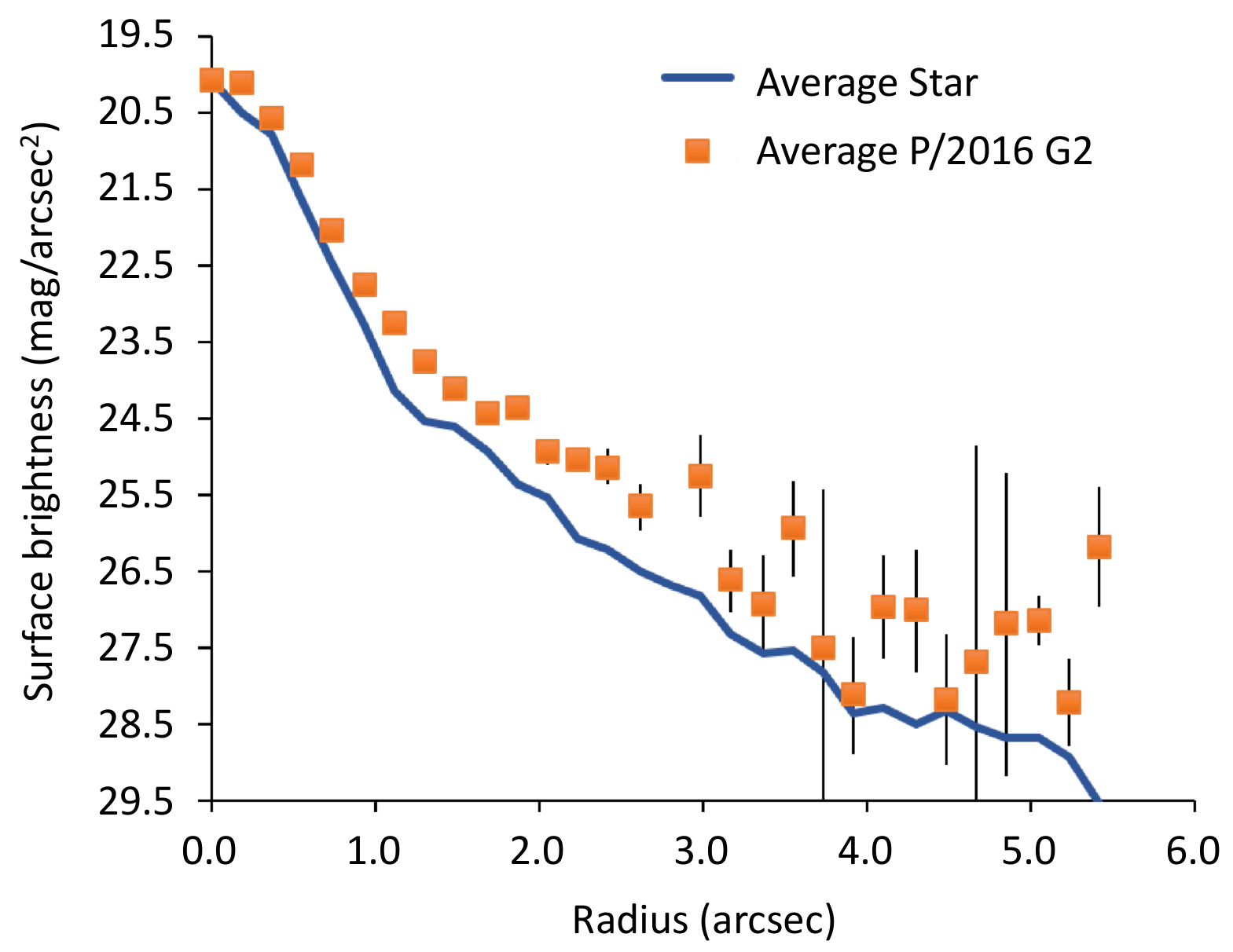}
\caption{\small
Average surface brightness profile of P/2016 G1 compared to the average profile for field stars as seen in the data obtained with the Pan-STARRS1 telescope on 2016 March~7.  The profiles for the field stars have been normalised to the peak brightness for P/2016 G1.  Although the images of the comet appear stellar in the \B{images, its profiles indicate that the object is slightly --but significantly-- extended.}
}
\label{fig-phot}
\end{figure}
%-^---^---------------------------------------------

\section{Analysis \& Results}
\label{sec:results}

\subsection{Surface Brightness Profiles}
\label{sec:SB}

The pre-\B{dis}covery data on \PG\ with the PS1 telescope on 2016 Mar. 7 consist of four images, showing the object with a stellar morphology. Only two of the frames had the object clear of the chip gap boundary. The surface brightness profile was computed in each of these frames for \PG\ and for three nearby field stars in order to assess whether there was any dust surrounding the nucleus.  We compute the radial flux profile using the median per-pixel flux in circular annuli around the object, and convert the flux to surface brightness using the zeropoint in the PS1 headers.  Errors are obtained using bootstrap re-sampling within each annulus. Figure~\ref{fig-phot} presents the average normalised surface brightness profiles of both \PG\ and the field stars, and shows that \PG\ is more extended. During each exposure, the object moved about 0$\farcs$3, which is significantly less than the seeing, so that trailing does not cause the flux excess seen in \PG. We conclude that the object was already surrounded by dust on 2016 Mar.~7.

\subsection{Description and nomenclature}\label{nomenclature}

Figure \ref{fig:images} shows a series of images of the object over time.
The central structure of the object is composed of 3 main concentrations of dust forming an inverted ``C'' on the images \B{(hereafter referred to as the three clumps)}, and a much fainter concentration on the eastern side, leaving a central gap. This structure is growing with time, preserving its appearance. The central structure is bathed in a diffuse, elongated coma, that extends into a dust tail westward from the central structure. The orientation of the coma and the tail is rotating southward as the viewing geometry evolves.

Below the central structure, a small, narrow, well-defined arc extends toward the South-South-East. It is clearly seen from Apr.~3 until May~12 and is still detected on May~31. Directly eastward of the central structure, but not pointing toward its centre, a faint linear feature is present. It can also be seen until May~12. The arc and the linear features are marked on Fig.~\ref{fig:images}C.

\subsection{Main Head Dust Feature\label{head}}
\label{sec:head}

%--FIG-4----------------------------------------------
\begin{figure}[ht!]
\includegraphics[width=8.5cm]{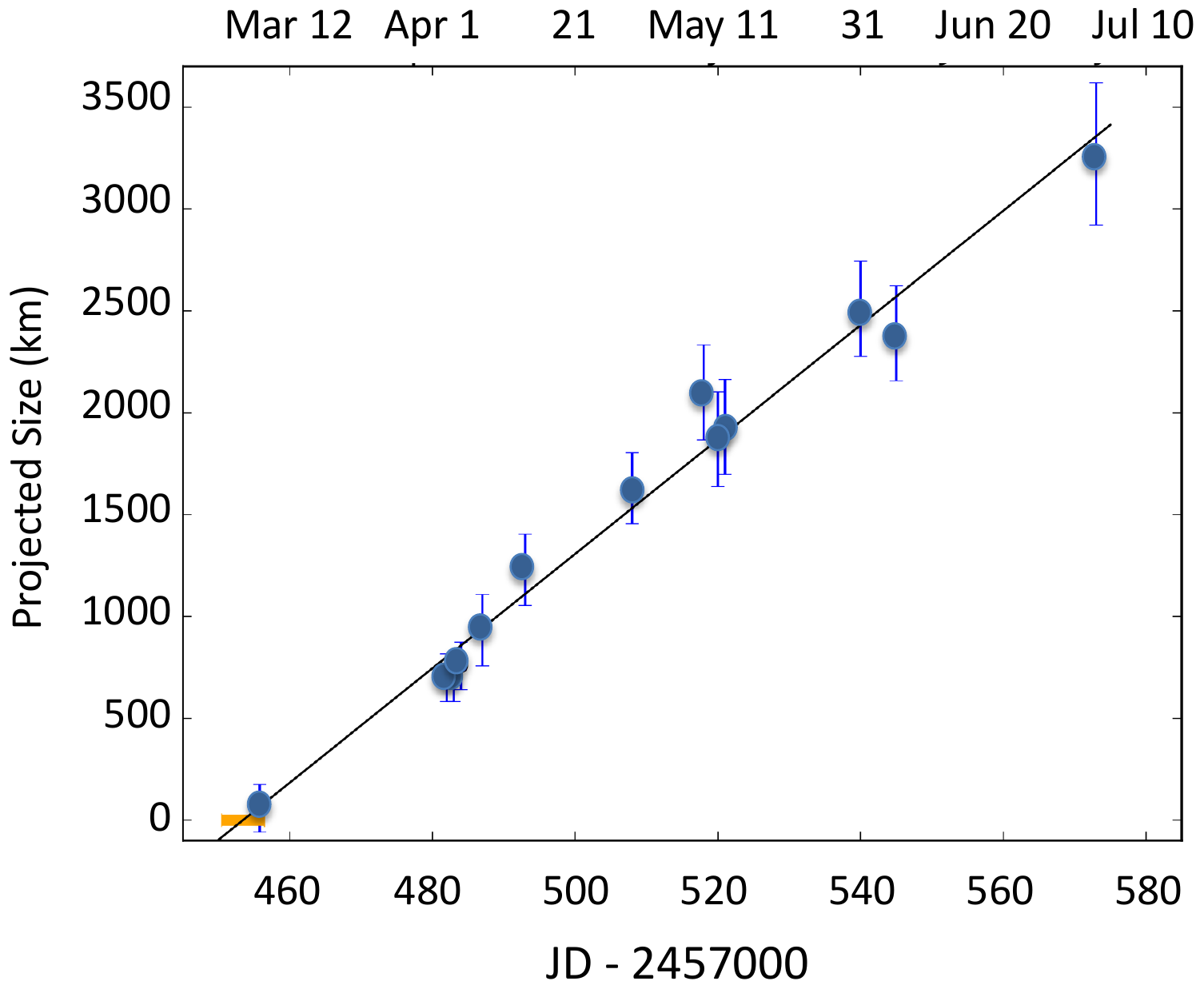}
\caption{\small
Size of the main head structure as a function of time. The line and the horizontal error bar indicate the linear expansion fitted to the data and the error on the origin, 2016 Mar. 6 $\pm$ 3~days.
}
\label{fig:size}
\end{figure}
%-----------------------------------------------------

The Earth crossed the comet's orbital plane on 2016 Apr. 13. The images taken around that date indicate that the complex head structure and the central coma extends above and below the orbital plane, whose orientation corresponds on the images to that of the tail feature (PA$=274^\circ$). The grains forming the head structure must, therefore, have been ejected with a non-zero velocity, with a non-zero component perpendicular to the orbital plane.

The size of the central structure was estimated measuring the distance between the peak of the North and South concentrations. When the peak is not well defined, the centre of the concentration was used instead. These angular distances were converted to a linear distance in the plane of sky accounting for the geocentric distance (and reported in Fig.~\ref{fig:size}). In the Mar.~7 image, \B{although the object is broader than stellar,}the central structure is not resolved yet. The diameter of the seeing disk is therefore used as an upper limit. The size of the structure grew linearly with time. A linear regression indicates that the growth would have started on 2016 Mar.~6 (JD=2457454) $\pm$ 3 days, and is expanding with an average (plane of sky) velocity of 0.32$\pm$0.02 m s$^{-1}$ (i.e. drifting from the centre of the original object at half that speed). The Mar.~7 image would therefore have been obtained very soon after the onset of the activity, \B{the observed broadening of the image could be caused by the cloud of fine dust expanding more rapidly than the larger fragments.}

\subsection{Finson-Probstein Dust Modelling\label{FPmodel}}

The positions of the Earth, Comet and Sun in late May and in June show the dust features of the tail with a geometry very favourable for a Finson-Probstein \citep[FP][]{finson68, farnham96} dust dynamical analysis.  FP modelling calculates the trajectories of an ensemble of dust grains of different sizes, parameterised by $\beta$ (ratio between the radiation pressure and the solar gravity) ejected from the surface of the nucleus at different times, $\tau$, as acted upon by solar gravity and solar radiation pressure.  We used the FP approach to analyse the pattern of synchrones (loci of the particles emitted at the same time) and syndynes (curves joining particles with the same $\beta$) and the optical appearance of the comet's dust environment.

% %-V-FIG-V----------------------------------------------
\begin{figure}[ht!]
%\plotone{Fig-fp.pdf}
\begin{center}
\includegraphics[width=8.5cm]{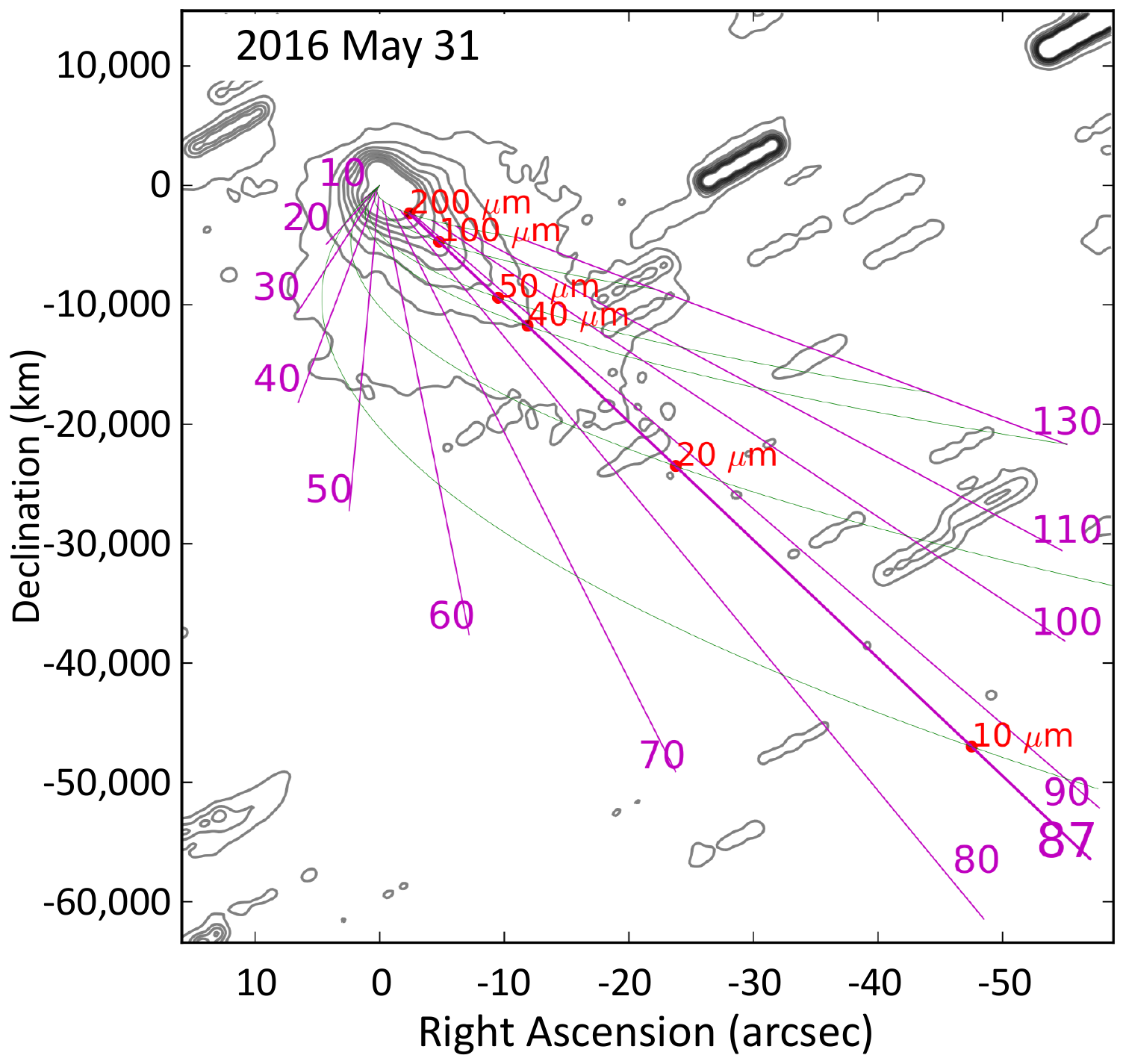}
\includegraphics[width=8.5cm]{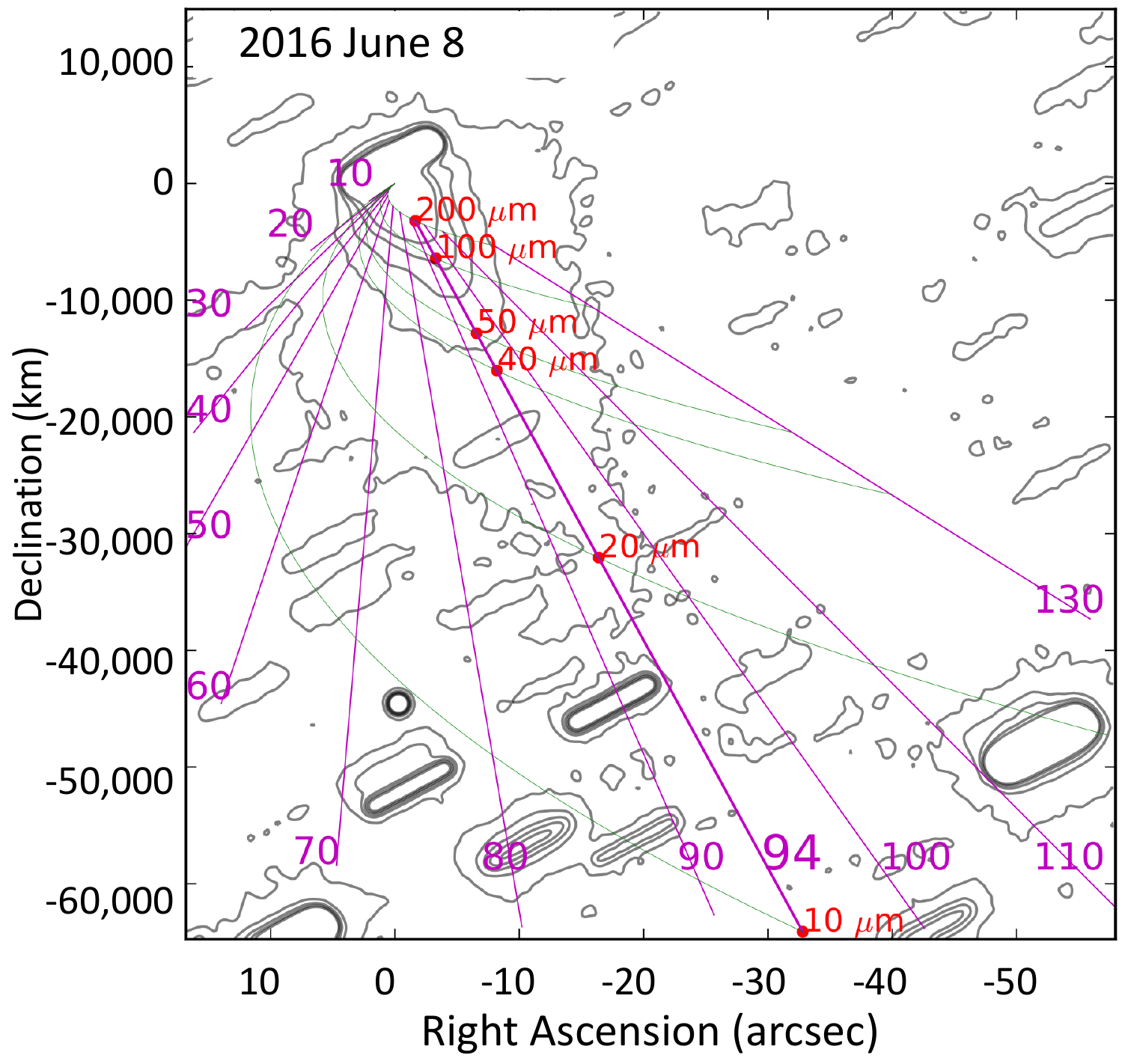}
\end{center}
\caption{\small Syndynes (in green) and synchrones (in purple, labelled in days before the observation date) for P/2016 G1 on 2016 May 31 and 2016 Jun. 8. The thicker synchrone marks the brightest peak in the dust profile, and the corresponding grain radii are marked in red. This fits with the disruption taking place around March 6. 
}
\label{fig:fp}
\end{figure}
%-------^----------------------------------------------

The main tail-like feature has a fairly sharp profile in azimuth as shown in Fig.~\ref{fig:fp}.  The position angle of the peak was measured on the images; it is marked as a thick line on the FP plots.  The epoch of the corresponding synchrone was obtained by comparing the position angles of synchrones generated with a step of 1 day.  The error on that epoch resulting from the measurement uncertainty is $\pm$3 days.  The azimuthal profile of the tail is very roughly gaussian; converting the FWHM (measured using the same procedure as above), this also results in a ``$\sigma$'' of $\pm$3 days.  The broadening of the tail is caused in part by the seeing (negligible far from the nucleus), by the dispersion in emission velocity (which is neglected in the zero-velocity FP method) and by the duration of the dust emission. The broadening caused by the emission velocity can be estimated from the images obtained when the Earth was close to the orbital plane (when the synchrones and syndynes degenerate into a single line), around Apr.~13 (see Fig.~\ref{fig:images}): \B{the width of the tail on Apr.~14 is similar to that seen on Apr.~4--8, when the orbital plane was observed at an angle.} We conclude that the broadening of the tail is dominated by the velocity dispersion of the grains \B{(which occurs in all directions, including perpendicular to the orbital plane), rather than purely by the radiation pressure (which occurs only within the orbital plane)}.

The tail-like feature is therefore compatible with a burst of dust emission centred on 2016 Mar.~4$\pm$3 days for the image from 2016 May 31, and 2016 Mar.~7$\pm$3 days for the image from 2016 Jun. 8. The emission profile is compatible with a short, impulsive burst smeared out by a distribution of initial velocities of the particles with $v < 1$\ms\ in random directions, or a longer burst (increasing then decreasing over a few days) with no initial velocity. Furthermore, dust is present in the areas of the image not covered by the syndynes and synchrones, confirming that the dust must have been emitted with some initial velocity. We therefore favour the short, impulsive burst.

It is interesting that two independent measurements for the date of the onset of disintegration: the FP estimates and the central feature growth are in perfect agreement.

Assuming that the tail-like structure is dominated by dust emitted with zero velocity, the spread of the dust along the tail is caused only by the radiation pressure. Close to the head of the object, the actual velocity dispersion of the grains will smear this relation, but further away, $\beta$ (the ratio between the radiation pressure and the solar gravitation for the considered grain), $a$ (the grain radius) and $\rho$ (its density) are related by:
\begin{equation}
\beta = 5.740 \times 10^{-4} \times \frac{ Q_{\rm pr}}{ \rho a} .
\end{equation} 
\noindent 
$Q_{\rm pr}$ is a radiation pressure efficiency coefficient, in the 1--2 range for rocky and icy material. With this relation, we can estimate the characteristic grain sizes in the images: with $\rho = 3000$~kg~m$^{-3}$, and $Q = 1.05$, Eq. 1 gives $ a = 2\times 10^{-7} / \beta$~[m].  The loci of 10 to 200~$\mu$m grains are marked on Fig.~\ref{fig:fp}. Closer to the head than the 100$\mu$m mark, the blurring by the seeing and the non-zero velocity of the grains forming the head structure prevent FP from making meaningful estimates of the grain size beyond ``they are large'', {\it i.e.} in the mm range or even larger.

\subsection{Dynamical Model of the Arc and Linear Feature\label{HeadModel}}

% %-V-FIG-V----------------------------------------------
\begin{figure}
\includegraphics[width=8.5cm]{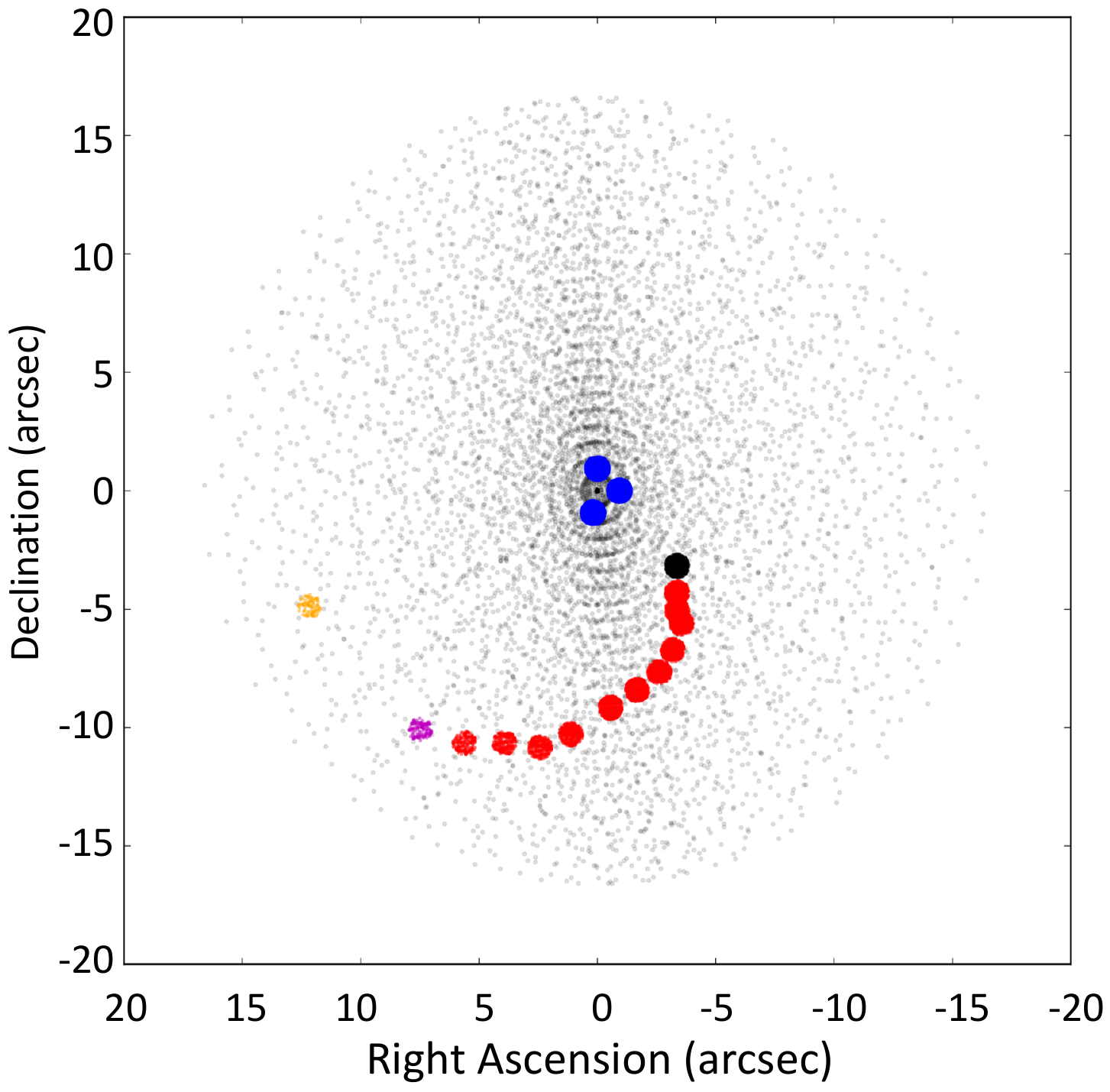}
\caption{\small
The positions of large particles emitted by the nucleus on 2016~Mar.~6, with velocities ranging from 0 to 3\ms\ observed on 2017~May~12 are marked with light grey dots in the plane of sky. The coordinates mark offsets from the position of the nucleus for that time. For clarity, only a subset of the particles is plotted. The particles matching the position of the 3 head features are marked in blue. Those matching the arc are marked in red (the ends of the feature are marked in magenta and black); the blob at the western start of the linear feature (Fig.~\ref{fig:images}) is marked in orange.
\label{fig:bubbleSky}
}
\end{figure}
%-------^----------------------------------------------
%-V-FIG-V----------------------------------------------
\begin{figure*}
\includegraphics[width=18.8cm]{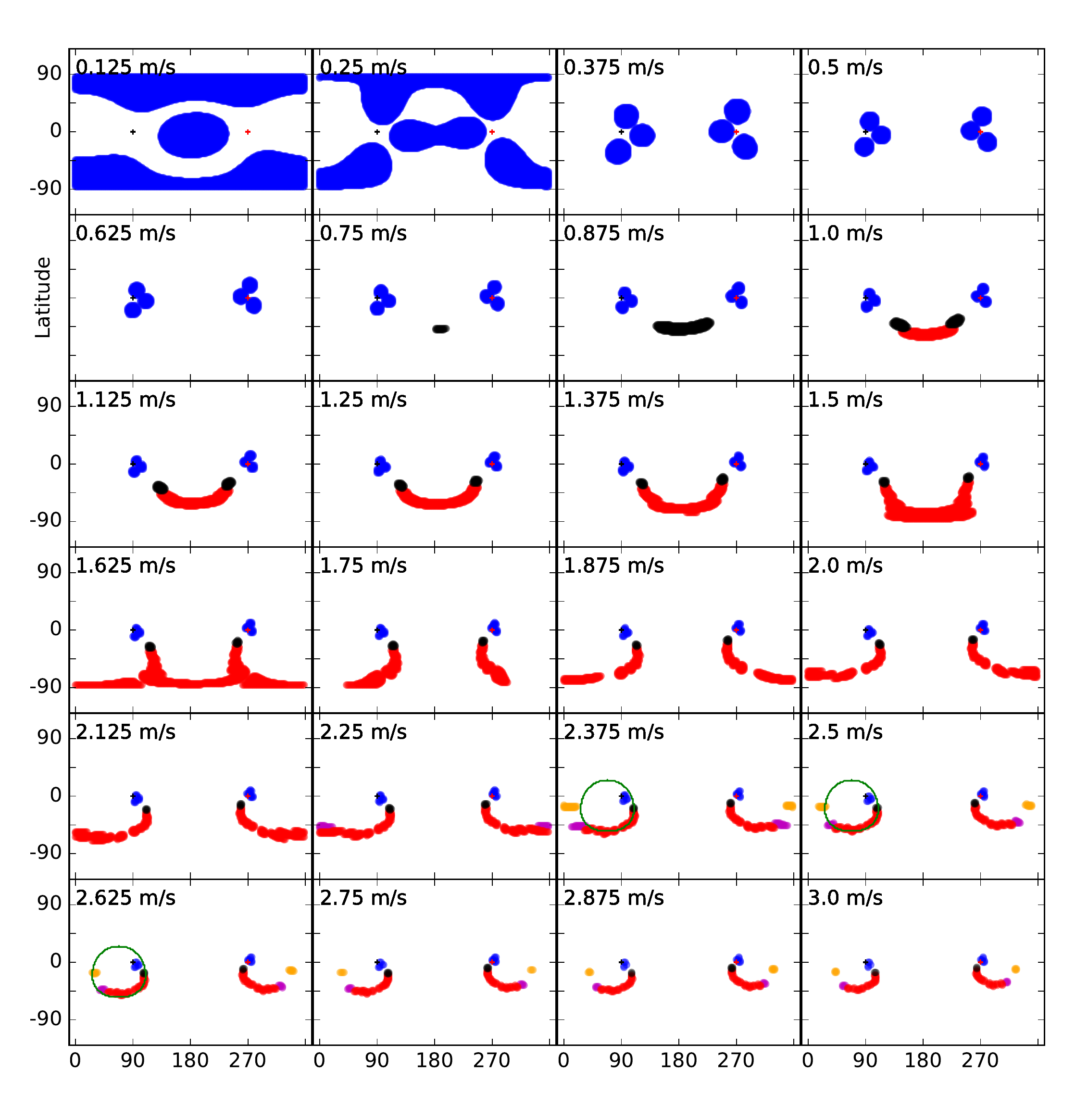}
\caption{\small
For a series of emission velocity, maps of the emission direction (in cometocentric longitudes and latitudes) of the large particles matching the features observed on May 12, using the same colour code as in Fig.~\ref{fig:bubbleSky}. For any velocity, a particle matching a feature always has a "mirror particle": one is emitted toward the Earth, the other one away from the Earth. On the 2.375, 2.5 and 2.625\ms\ panels, the position of a $40^\circ$ half-opening cone matching best the feature is represented. The direction directly toward Earth is marked by a black cross, away from Earth by a red cross, at $0^\circ$ latitude, and $90^\circ$ and $270^\circ$ longitude respectively. Re-projecting these maps onto concentric spheres, with the geometry of May~12, would lead to Fig.~\ref{fig:bubbleSky}.
}
\label{fig:bubbleMap}
\end{figure*}
%---^--------------------------------------------
%-V-FIG-V----------------------------------------------
\begin{figure}\begin{center}
\includegraphics[width=7.0cm]{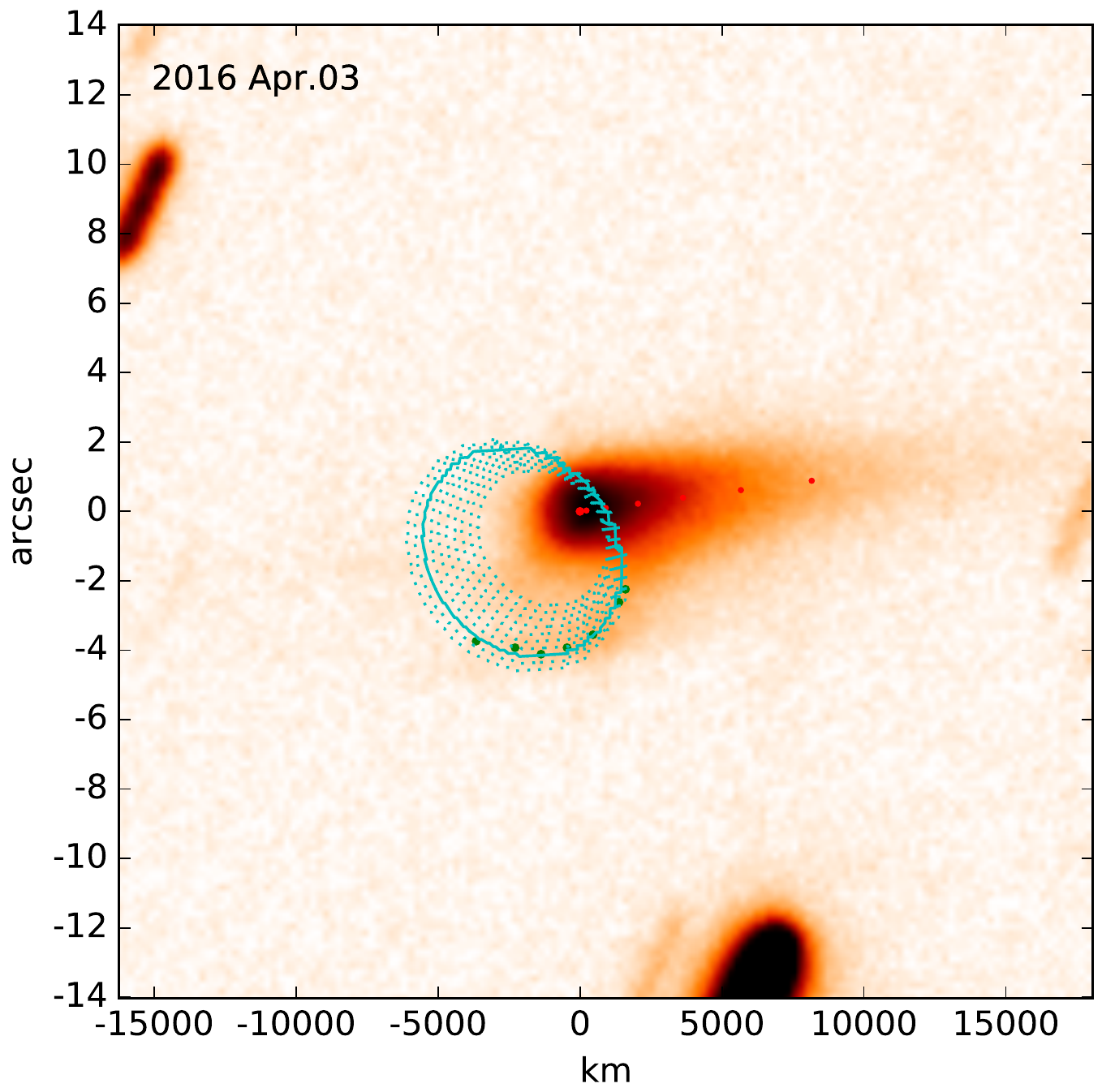}
\includegraphics[width=7.0cm]{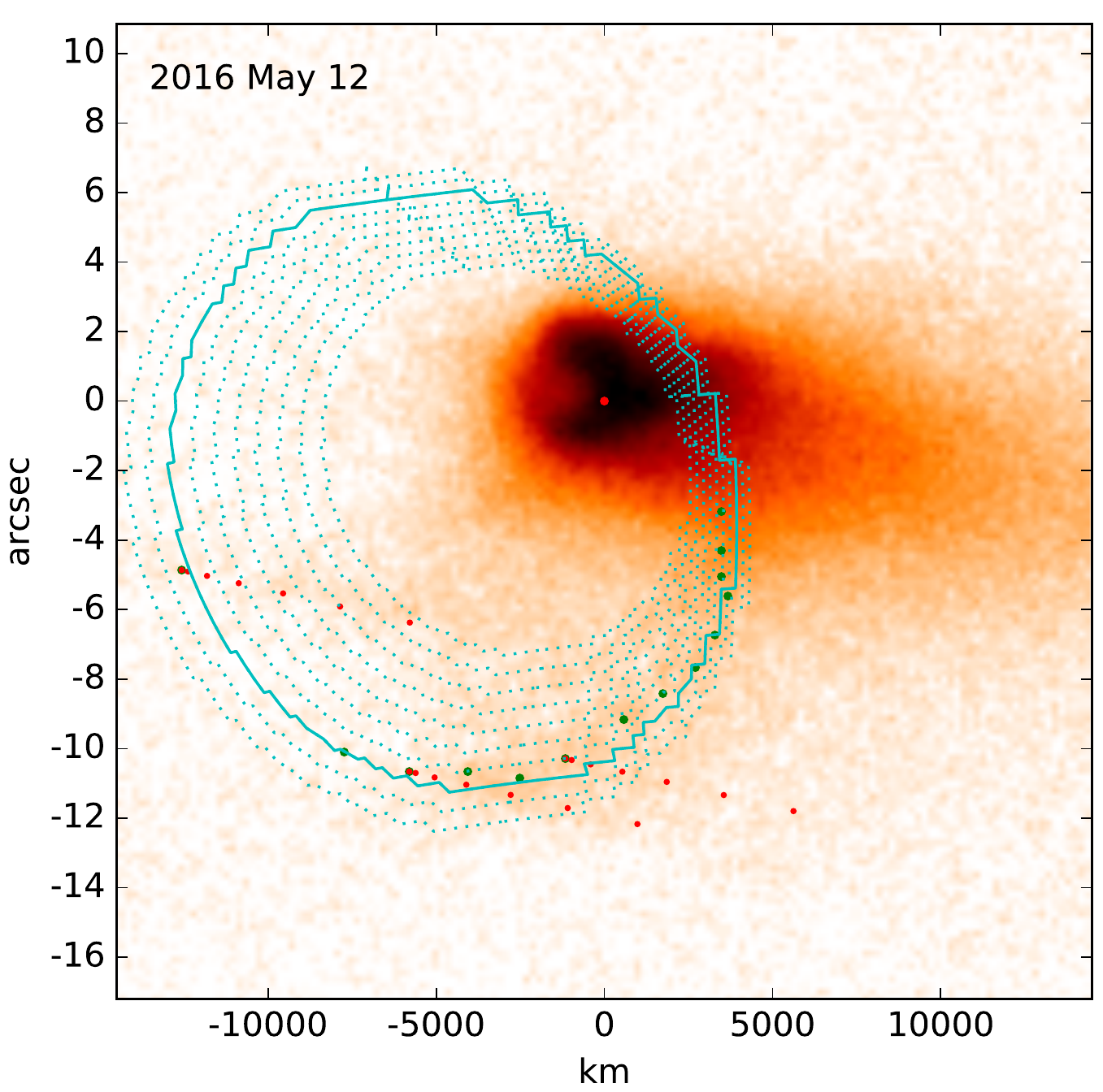}
\includegraphics[width=7.0cm]{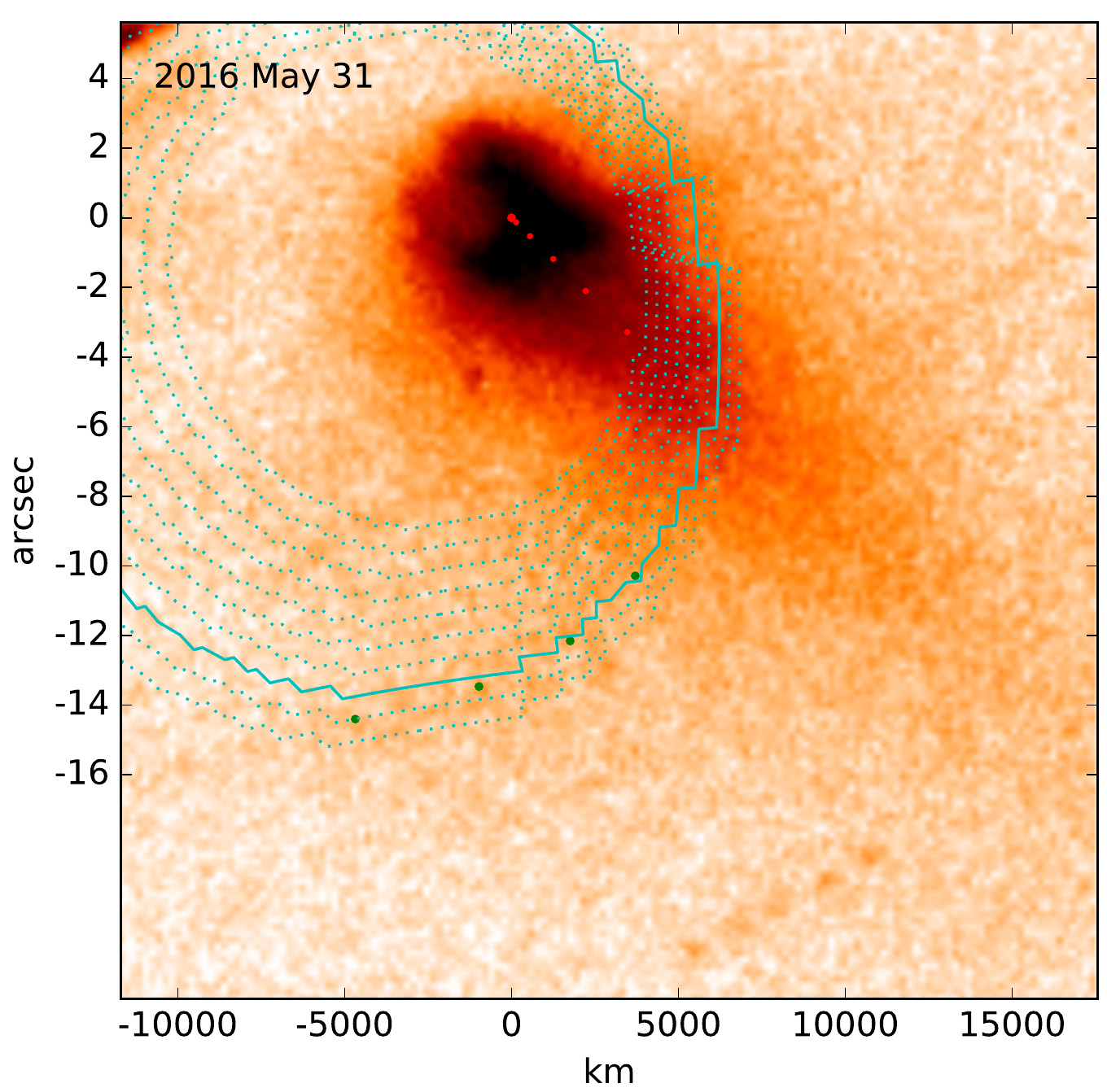}
\end{center}
\vspace{-0.2cm}
\caption{\small
P/2016 G1 on 2016~Apr.~3, May~12 and 31. The green dots mark clumps on the arc feature. The small red dots correspond to the synchrone trajectories of small dust (submitted to radiation pressure) emitted from some of the clumps on Mar.~6. The position of large particles emitted on Mar.~6 on a cone with half-opening 40$^{\circ}$ and centred on (long., lat.) = (68$\deg$,-15$\deg$) are over-plotted for velocities ranging from 1.625 to 2.750\ms. The solid line corresponds to 2.5\ms. 
\label{fig:coneImages}
}
\end{figure}
%---^--------------------------------------------

In order to further investigate the structure of the head, we simulated the trajectory of large particles (i.e. not influenced by radiation pressure, $\beta = 0$), emitted from the nucleus' position on 2016~Mar.~6 with velocity values ranging from 0 to 3~m~s$^{-1}$, and directions covering the whole sphere every 2.5$^{\circ}$ in longitude and latitude \B{(cometocentric coordinates, $x$ and $y$ in the orbital plane), $y$ pointing toward the Earth)}. The position of each particle was computed for 2016~Apr.~3, May~12, and May 31. The particles that were located at the position of the 3~fragment clumps, the  South-East arc and the start of the linear feature are identified, as illustrated in Fig.~\ref{fig:bubbleSky}. Figure~\ref{fig:bubbleMap} shows the maps of the particles matching the features (in cometocentric longitude and latitude at the time of the emission), for a set of emission velocities.

{\bf The three main fragment clumps} can be reproduced by particles emitted at any velocity between 0.25 and 3\ms. However, for higher velocities, they must be emitted almost exactly toward or away from Earth (so that the foreshortening keeps them at their position near the origin), which is unlikely. It is more likely that they were emitted at a random angle, i.e. at velocities $<$ 1\ms.

{\bf The South-East Arc} cannot have been emitted at velocities $<$ 2.25\ms: part of the arc is not reproduced (some of the dots are missing). For velocities in the 2.25 -- 3\ms\ range, the complete feature can be reproduced by particles emitted at the same velocity. A cone with a half-opening of $40^\circ$, whose axis points toward cometocentric coordinates (long., lat.) = (68$\deg$,-15$\deg$), is found to reproduce well the arc feature and the start of the linear feature, with a velocity of 2.5\ms: the arc and the clump at the head of the linear features are on a ring corresponding to the 2.5\ms slice of this same cone. 

In order to test the validity of this interpretation, the position of the particles was also computed for Apr. 3 and May 31, when the viewing geometry was very different. As illustrated in Fig.~\ref{fig:coneImages}, the same ring of dust matches well the 3D position and evolution of the arc and the head of the linear feature.

Adjusting the position and opening of the cone (by eye) gives an estimate of the uncertainty on these parameters: $\pm 2.5^\circ$ in longitude and latitude, and $\pm 5^\circ$ in opening. Changing the opening of the cone also slightly modifies the velocities of the particles matching the features. The date of the initial burst was also changed by 5, 10, 15~days. No effect is noticeable for 5~day; a slight change of orientation and velocity is required for the 10 and 15~day shifts. This therefore does not set additional constraints on the time of the emission.

Using the FP formalism (Section~\ref{FPmodel}), the trajectory of small dust grains emitted together with the larger grains was calculated accounting for the radiation pressure. The smaller grains form a line (synchrone) starting at the large particle, and drifting away at an angle set by the emission date (while the large particles move on a straight line in the image plane, from the nucleus toward their observed position, the small particles start tangentially to that line, and drift on a parabolic particle toward their observed position).  \B{The eastern linear feature matches well the position of small grains emitted at the same time, in the same direction and at the same velocity as the clump at the head of the linear feature. They are highlighted with red dots in the May~12 panel of Fig.~\ref{fig:coneImages}.}
Similarly, various clumps in the arc feature have similar ``tails'', which together cause an enhancement of surface brightness on the western side of the arc \B{(two of them are highlighted with red dots in the May~12 panel of Fig.~\ref{fig:coneImages})}. As the position angle of these synchrone lines is set only by the time of the emission, the excellent match between the eastern linear feature and the synchrone for a Mar.~4 emission gives a very strong support to the hypothesis that it corresponds to small grains emitted simultaneously and in the same direction as the big grains, and further support the Mar.~4 emission date. It also rules out that the linear feature would have been caused by a secondary disruption of a large grain at a later date: the position angle of the corresponding synchrone would have been different.

Overall, the southern arc and the head of the eastern linear feature are large particles partially populating a ring. The arc covers at least a quarter of the ring, possibly more, as its Northern part is lost in the main coma. A single clump of large particles on the ring forms the tip of the eastern linear feature. This ring grows in size with time, as the particles move away from the emission spot at $(2.5 \pm 0.1)$\ms, describing a cone with an half-opening of 40$^{\circ}$. 

It is possible that the sections of the ring that appear devoid of dust were originally populated by small grains: solar radiation pressure would have pushed them away from the field of view.

The ring corresponds to a very narrow velocity distribution: in particular, the ``walls'' of the cone are not populated by dust grains, only a (partial) ring. 

We tried and failed to reproduce the arc and the eastern linear feature using dust grains emitted along a great circle in cometocentric coordinates, as would have been produced by an equatorial centrifugal ejection. We also tried and failed to reproduce the observed features using grains with a broader range of velocities: a velocity distribution around 2.5~\ms, narrower than 0.125~\ms, gives the best results.

\subsection{The Moreno Model}\label{sec:moreno}

\citet{moreno2016} obtained data for P/2016 G1 on three nights between 2016 Apr.~20 and Jun.~8 and used Monte Carlo techniques to model the dust. They obtained two additional epochs with the Hubble Space Telescope on 2016 Jun.~28 and Jul.~11 \citep{moreno2017}. 

Comparing the original dataset to model images generated using Monte-Carlo emitted particles following a FP-like formalism, they infer that the dust ejection began around 2016 February 10$^{+10}_{-30}$, and decreased with a half-Gaussian function with a half-width at half maximum (HWHM) of $24^{+10}_{-7}$~d, corresponding to an emission of 1.7$\times$10$^7$ kg of dust, of which some was preferentially ejected in the westward direction, consistent with an impact aligned with the sun-asteroid vector. They inferred very low dust velocities with average speeds around \B{0.8\ms}\ and grain sizes between 1$\mu$m to 1~cm in radius\B{(the velocities in the original paper \citet{moreno2016} had a typo, corrected in an erratum \citet{moreno2019})}. The more recent HST data are compatible with their original model. 

Their analysis was focused on what we call here the central structure with the 3 main concentrations forming the inverted C on the images and the central dust coma. From an FP analysis similar to that of Section~\ref{FPmodel}, they set the interval for the beginning of activity in their model from 2016 Jan.~31 to Apr.~1. 

Our estimate of the date of the peak of dust emission relies on two simple and independent measurements:
$i)$ the linear growth of the separation between the condensations in the central structure, which agrees with the barely resolved image obtained on Mar.~7 (Section~\ref{sec:SB}), and
$ii)$ the position angle of the sharp cusp in the isophotes far from the origin, in Fig.~\ref{fig:fp}. Measuring this angle on Moreno's Fig.~2, we get a very good agreement with our estimate. This makes the simple assumption that far from the origin, the effect of the radiation pressure dominates over an initial velocity. Moreno et al. run their Monte-Carlo multi-parameter model to refine the combination of initial velocity and radiation pressure, leading to a difference in emission time of about a month. 

Their analysis does not mention the southern arc, nor the eastern linear features, whose geometry we find to strongly and independently confirm the emission date provided by the orientation of the main tail and the growth of the central structure, and to also support a very brief emission. This is further confirmed by the very narrow eastern linear feature, that points precisely at an impulsive emission on again the same date, Mar.6. Because of the convergence of these independent arguments, because they are based on much wider features than those Moreno et al. worked on, and because of the simplicity of the underlying assumptions, we are confident that the disruption occurred on 2016 Mar. 6$\pm$3~days.

In terms of duration, the analysis of Moreno et al. suggests that ``[an] impact would have induced a partial destruction of the asteroid,'' causing the anisotropic part of the dust emission, ``with dust grains being emitted to space nearly isotropically while the body is being torn apart,'' that this isotropic emission would have had a duration of $24^{+10}_{-7}$~d HWHM. Our data directly support the impact-induced destruction of the asteroid,  the arc and linear feature being the signature of that impact. Concerning the broad dust feature on which Moreno et al. focus their study, our interpretation is that the broadening of the tail matches the dust velocity dispersion, and that therefore it is compatible with a very short duration event, although we cannot rule out (nor constrain) a longer duration for the final crumbling of the asteroid.

\subsection{Mass of the debris cloud}

%-V-FIG-V----------------------------------------------
\begin{figure}
\includegraphics[width=8.7cm]{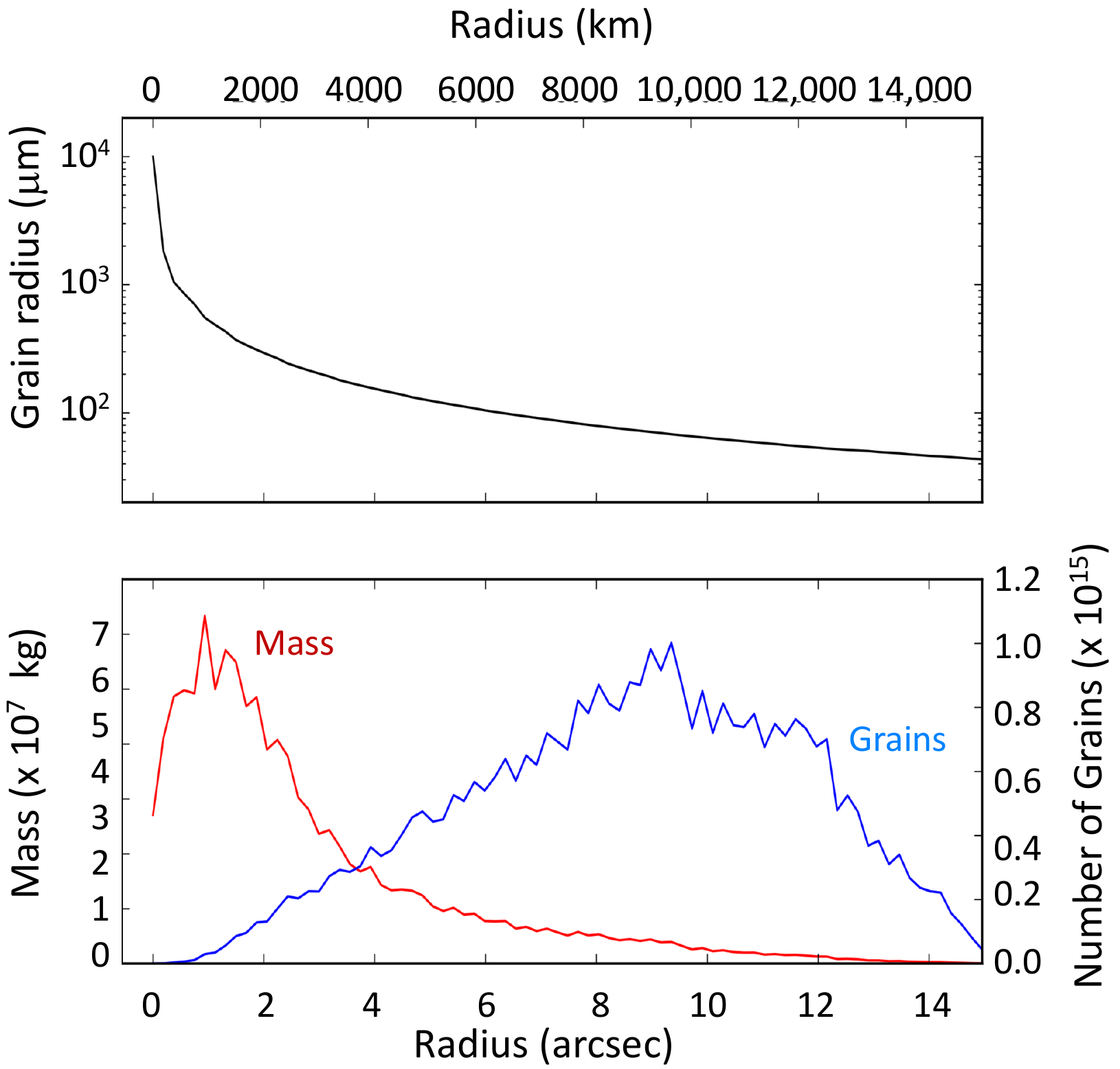}
\caption{\small Top: radius of the dust grains as a function of the distance to the centre of the object, using the FP scaling (see Sect.~\ref{FPmodel} and Fig.~\ref{fig:fp}) for May 31. Bottom: the flux in the May 31 image is integrated over concentric rings and converted into the number of particles (blue), and in mass (red)\B{, both in each radial bin}.
}
\label{fig:mass}
\end{figure}
%---^--------------------------------------------

Because the exact grain size distribution present in each pixel of the image is not known, and because the central structure can contain very large particles (meter-sized or more), getting an evaluation of the mass of the object is difficult. This is made worse by the fact that larger particles contribute more to the mass, while smaller particles contribute more to the flux. Below comes an attempt to get a rough estimate of this mass, based on the May~31 image. We assume a dust density $\rho = 3000$~kg~m$^{-3}$, and an albedo $p=0.25$.

To get an order-of-magnitude estimate, one can consider a uniform size distribution across the whole object (a power law with a $-3$ index, and a cut-off at 1~cm, after \citet{moreno2016}). This leads to a total mass in the object of $1\times 10^7$~kg, in agreement with Moreno et al. who obtained an ejected mass of $2\times 10^7$~kg with their detailed model. 

As we consider that the dust release event was short, the dust distribution (far from the position of the nucleus) is dominated by the radiation pressure: smaller grains are pushed further away than larger ones. While this estimate is also a simplification, it takes into account the fact that the dust is not uniformly distributed. The grain size is set using the scaling between size and distance from Fig.~\ref{fig:fp}. The image is then integrated on concentric rings, and the flux is converted in dust surface, then in the number of grain, then in mass, as illustrated in Fig.~\ref{fig:mass}. This results in a total mass for the object $M_N = 1.1\times 10^9$~kg. This estimate is still a lower limit, as larger clumps could hide in the central concentration. The mass corresponds to a spherical object with a radius $R_N \sim 44$~m, well below the limit of detection of the pre-discovery images ($R_N<200$ m).

We don't know whether the object was {\em completely} disrupted. Using their high-resolution, deep HST images, \citet{moreno2017} obtained an upper limit of 30~m to the fragments. The mass potentially contained in even a few of these large fragments would completely dominate the mass of the dust. In what follows we consider as a lower limit the mass of the dust (and the corresponding radius $R_N > 44$~m), and the upper limit from the pre-discovery constraints ($R_N < 200$~m). Most of the mass being located in the central concentration, which we observed expanding at $v=0.16$\ms. The corresponding kinetic energy is $K=1.4\times 10^7$ and $1.3 \times 10^9$~J, respectively. In order to reach this asymptotic velocity, the grains and fragments had to overcome the (small) gravity potential of the object, with an escape velocity $v_e = \sqrt{\frac{2GM}{r_T}} = 0.06$\ms\ and $0.26$\ms, respectively. This implies that the disruption of the object required a total energy at least $E = \frac{1}{2} M_N (v^2 + v_e^2) = 1.6 \times 10^7$ and $4.7 \times 10^9$~J, respectively. These are clearly lower limits that consider a pure rubble pile, i.e. without and internal cohesion forces.

Considering that this energy was provided by an impactor, and that collision happened at the median impact velocity in the main belt $v_i= 5\,000$\ms, this requires a tiny impactor of 1~kg, with a radius of 5~cm for the lower limit on the object size, and 370~kg with a radius of 30~cm for the larger object.

\subsection{Summary of the analysis}\label{sec:analysissummary}

Overall, our results are based on various independent measurements and separate simple interpretation pointing toward the following scenario: 
\begin{itemize}
\item No image from before 2016 Mar. shows the object. These non-detections lead to a radius $R_N< 0.2$~km.
\item The onset of activity was on 2016 Mar. 6. Backward extrapolation of the central structure gives an uncertainty of $\pm$3 days but the object was found active on Mar.~7. This Mar. 6 date is confirmed by the following independent measurements: orientation of the azimuthal peak of the main tail; position and evolution of the southern arc and of the head of the linear feature; orientation of the eastern linear feature.
\item The dust emission event had a very brief component (indicated by the narrow eastern linear feature and the narrow ring of large particles populating the southern arc), which can possibly have been followed by a longer emission of dust from the central feature.
\item The central component of the object is composed of large dust grains moving at low velocity ($0.16\pm0.01$\ms\ in the plane of sky) from the original centre. It is surrounded by a cloud of smaller particles, also emitted at similar low speed, and being pushed away by radiation pressure. The visible tail is populated by particles in the 10--200~$\mu$m range. Larger particles are certainly present, but confused into the central component; smaller particles drifted away from the field of view. \B{The broadening of the Mar.~7 image could be caused by the hyper-velocity material released during the impact itself, while the slower material was released during the follow-up disruption}.
\item During the brief emission event, a group of particles was emitted on a 
cone with a half-opening angle of $40\pm5^\circ$, propagating at (2.5$\pm$0.1)\ms and forming part of a ring, with an arc covering over a quarter of the ring plus a clump opposite to the arc (forming the southern arc and the head of the eastern linear feature). The large particles were emitted together (in time, direction and velocity) with smaller particles, which are seen drifting from the ring under the radiation pressure. They form the eastern linear feature and a faint westward extension to the southern arc. These features could not have been reproduced by emission on a grand circle (matching, for instance, an equatorial ejection).
\end{itemize}

\section{Discussion: Impact \& Disruption\label{discussion}}

From the observations, a major disruption occurred on \PG\ on or around 2016 Mar. 6, disrupting --possibly destroying-- the object, and releasing a ring of large particles moving at $\sim 2.5$\ms, growing on a cone with a half-opening angle of $\sim 40^\circ$. We consider the impact of a meter-scale object hitting a 100-meter-scale asteroid at $\sim 5$~\kms. All the hyper-velocity ejecta which would be expanding at impact-scale speeds (in the \kms\ range) have moved degrees away from the position of the object at the time of the first deep images (about one month after the impact). Radiation pressure having further dispersed these grains, they are lost beyond the field-of-view of the observations. 

All that is left in the field-of-view are the spalls and large dis-aggregated fragments of the objects that came off well after the impact, in the final stages of the formation of the crater and after, together with the dust grains released.
\B{Crater formation simulations (numerical and laboratory) show that asymmetries in the filling of the ejection cone are common; this would explain why the ring is not complete.}

These grains ($>10\mu$m) are either the small end of a power-law size distribution of the fragments, and/or the regolith that was dragged along with bigger fragments. The bulk of the impactor's energy is transferred into the bulk of the body, causing its disruption into fragments that drift apart at very low velocities.

Hydrodynamic simulations of impacts can provide information on the remnants, in particular on their size distribution \citep[see][for a systematic study of various parameters]{durda2007} and velocities. Simulations of catastrophic impacts\footnote{
%https://www.dropbox.com/s/ryyb93tj8v0jjxm/gv13_movie.mov,
URLs: https://youtu.be/pqaqbfbevoo \\
https://youtu.be/e4HCTcQ-IWA} show that the plate-shaped spall fragments, corresponding to the final moments of the crater formation, are ejected with different velocities and over a broad range of directions. However, the crater formation process also produces wedge-shaped fragments, bounded by the propagating radial cracks. These have a narrow distribution of slow velocity, in the \ms\ range, and are axially isotropic about the point of impact. These large wedges would become visible as they break apart and their surface area increase. Based on these qualitative arguments, we suggest that the feature we interpret as a ring of fragments corresponds to the left-over of these large wedge-shaped fragments. Hydrodynamics simulations, beyond the scope of this paper, will be used to quantitatively verify the plausibility of this hypothesis. Interestingly, the velocity of ejection of these wedge-shaped fragments is directly linked to the tensile strength of the object. The tensile strength is one of the key structural characteristics of an asteroid, but it is normally not accessible remotely. In the case of a rotationally disrupted asteroid, the study of the fragmentation can cast some light on the internal strength, but with some uncertainties caused by the unknown initial spin of the object \citep[see][for an application to rotationally disrupted P/2013~R3]{hirabayashi14}. 
A result of the ongoing simulation will, therefore, be the first direct remote estimate of the tensile strength of an asteroid, a measurement that would have implication for the mitigation measures being investigated for planetary defence initiatives.

\begin{acknowledgements}
We are very grateful for the detailed review and comments provided by the referee, Fernando Moreno. KJM, JK, and JVK acknowledge support through an award from the National Science Foundation AST1413736.  RJW acknowledges support by NASA under grant NNX14AM74G. This paper is based on observations obtained with MegaPrime/MegaCam, a joint project of CFHT and CEA/DAPNIA, at the Canada-France-Hawaii Telescope (CFHT) which is operated by the National Research Council (NRC) of Canada, the Institut National des Sciences de l'Univers of the Centre National de la Recherche Scientifique of France, and the University of Hawaii. Data were acquired using the PS1 System operated by the PS1 Science Consortium (PS1SC) and its member institutions. We thank the staff of IAO, Hanle and CREST, Hosakote, that made these observations possible. IAO and CREST are operated by the Indian Institute for Astrophysics, Bangalore. The Pan-STARRS1 Surveys (PS1) have been made possible by contributions from PS1SC member Institutions and NASA through Grant NNX08AR22G, the NSF under Grant No. AST-123886, the Univ. of MD, and Eotvos Lorand Univ. We would like to thank Detlef Koschny (ESA) and Mikael Granvik for their helpful discussion on impact cratering mechanics and disruption. This research was supported by the Munich Institute for Astro- and Particle Physics (MIAPP) of the DFG cluster of excellence ``Origin and Structure of the Universe''.
\end{acknowledgements}

\bibliographystyle{aa} % style aa.bst
\bibliography{g1.bib} % your references Yourfile.bib

\begin{appendix}
\input table_sideways

\end{appendix}

\end{document}

%% file: table_sideways.tex
\begin{sidewaystable*}
\caption{Observations\label{tab-data}}
\setlength\columnsep{2pt} 
\resizebox{\textwidth}{!}{%magic command to automatically resize table to fill
\begin{tabular}{cccccccccccccccc}
\hline\hline             
\multicolumn{10}{l}{Archive data}\\
\hline
{UT Date} & {Tel$^a$}
 & {Mid-JD$^b$}
 & {Filt}
 & {\#$^c$}
 & {Exp$^d$}
 & {$r$$^e$}
 & {$\Delta$$^e$}
 & {$\alpha$$^e$}
 & {PA$_{-\odot}$$^f$}
 & {PA$_{-v}$$^f$}
 & {TA$^f$}
 & {$m$$^h$}
 & {$R_{4\%}$$^i$}
 & {$R_{25\%}$$^i$}
 & {} \\
 \hline
2000-10-20 & INT  & 1838.45737 & z &    &      & 2.163 & 1.351 & 19.38  &  74.6 & 255.9 &   46.9 & $>$21.03 & 0.74 & 0.30 & \\
2007-02-16 & CFHT & 4147.93142 & i &    &      & 3.043 & 2.111 &  7.44  & 319.2 & 284.2 & -153.7 & $>$21.53 & 1.03 & 0.41 & \\ 
2011-02-23 & PS1  & 5615.91806 & $r_{P1}$,$i_{P1}$ & $2$,$2$ & 170 & 3.084 & 2.111 &  3.99  &  25.4 & 284.3 & -161.5 & $>$20.5  & 1.58 & 0.63 & \\
2011-05-25 & PS1  & 5706.78070 & $w_{P1}$          &       4 & 180 & 2.993 & 2.899 & 19.71  & 110.6 & 282.6 & -146.2 & $>$21.5  & 1.78 & 0.71 & \\
2012-06-09 & PS1  & 6087.93886 & $i_{P1}$          &       2 &  90 & 2.232 & 1.242 &  7.73  & 159.6 & 268.5 &  -59.6 & $>$22    & 0.36 & 0.14 & \\
2012-06-13 & PS1  & 6091.86993 & $w_{P1}$          &       6 & 270 & 2.224 & 1.242 &  8.97  & 147.7 & 268.3 &  -58.3 & $>$22    & 0.37 & 0.15 & \\
2013-10-28 & PS1  & 6594.13022 & $z_{P1}$          &       2 &  60 & 2.599 & 1.734 & 13.22  & 274.8 & 269.3 &  103.8 & $>$21    & 1.03 & 0.41 & \\
2013-11-04 & PS1  & 6601.05904 & $g_{P1}$,$r_{P1}$ & $2$,$2$ & 166 & 2.615 & 1.703 & 10.61  & 279.9 & 268.7 &  105.4 & $>$21    & 0.97 & 0.39 & \\
2013-11-28 & PS1  & 6624.92394 & $w_{P1}$          &       4 & 180 & 2.667 & 1.693 &  4.23  &   0.9 & 267.3 &  110.6 & $>$22    & 0.55 & 0.22 & \\
2014-11-18 & PS1  & 6980.07870 & $w_{P1}$          &       4 & 180 & 3.123 & 2.994 & 18.45  & 289.3 & 283.0 &  173.8 & $>$22    & 1.49 & 0.59 & \\
2014-11-28 & PS1  & 6990.11418 & $w_{P1}$          &       4 & 180 & 3.125 & 2.853 & 18.28  & 291.4 & 282.9 &  175.4 & $>$22    & 1.41 & 0.56 & \\
2014-11-30 & PS1  & 6992.05683 & $w_{P1}$          &       4 & 180 & 3.126 & 2.825 & 18.19  & 291.8 & 282.8 &  175.7 & $>$22    & 1.40 & 0.56 & \\
2014-12-02 & PS1  & 6994.09527 & $w_{P1}$          &       4 & 180 & 3.126 & 2.797 & 18.09  & 292.2 & 282.8 &  176.1 & $>$22    & 1.38 & 0.55 & \\
2015-01-17 & PS1  & 7040.03093 & $w_{P1}$          &       2 &  90 & 3.126 & 2.263 & 10.16  & 311.2 & 283.0 & -176.7 & $>$22    & 0.96 & 0.39 & \\
2016-01-10 & PS1  & 7398.14409 & $w_{P1}$          &       4 & 180 & 2.699 & 2.904 & 19.79  & 287.9 & 277.1 & -113.9 & $>$23    & 0.81 & 0.32 &\\
2016-03-07 & PS1  & 7455.11400 & $w_{P1}$          &       3 & 135 & 2.575 & 2.028 & 20.89  & 282.8 & 273.9 & -101.2 & 18.99$\pm$0.06 & & & \\
\hline
\multicolumn{10}{l}{New data}\\
\hline

UT Date 
& Tel$^a$ 
& Mid-JD$^b$  
& Filt 
& \#$^c$ 
& Exp$^d$ 
& $r$$^e$ & $\Delta$$^e$ & $\alpha$$^e$ & PA$_{-\odot}$$^f$ & PA$_{-v}$$^f$
 & TA$^f$
 & $m_{10k}^j$  & $m_{7.5k}^j$  & $m_{5k}^j$  & $m_{2.5k}^j$ \\
\hline
2016-04-03 & CFHT & 7482.06559 & w &  3 &  540 & 2.513 & 1.678 & 15.400 & 278.2 & 274.0 &  -94.8 & 18.806$\pm$0.002 & 18.982$\pm$0.002 & 19.424$\pm$0.002 & 20.688$\pm$0.003\\
2016-04-04 & CFHT & 7483.04336 & w &  3 &  540 & 2.511 & 1.668 & 15.104 & 277.9 & 274.0 &  -94.5 & 18.793$\pm$0.002 & 18.975$\pm$0.002 & 19.457$\pm$0.002 & 20.758$\pm$0.003\\
2016-04-05 & CFHT & 7484.05387 & w &  3 &  540 & 2.509 & 1.657 & 14.792 & 277.6 & 274.1 &  -94.3 & 18.789$\pm$0.002 & 18.973$\pm$0.002 & 19.450$\pm$0.002 & 20.737$\pm$0.003\\
2016-04-08 & CFHT & 7487.04616 & w &  6 & 1080 & 2.502 & 1.626 & 13.827 & 276.6 & 274.1 &  -93.6 &    &   &   &  \\
2016-04-14 & CFHT & 7493.05300 & w &  5 &  600 & 2.488 & 1.569 & 11.712 & 273.9 & 274.3 &  -92.1 & 18.721$\pm$0.002 & 19.005$\pm$0.002 & 19.622$\pm$0.002 & 21.012$\pm$0.004\\
2016-04-29 & CFHT & 7507.96100 & w &  3 &  360 & 2.454 & 1.465 &  5.838 & 255.1 & 274.6 &  -88.3 & 18.556$\pm$0.002 & 18.945$\pm$0.003 & 19.642$\pm$0.003 & 21.078$\pm$0.006\\
2016-05-10 & CFHT & 7518.94939 & w &  3 &  360 & 2.429 & 1.425 &  3.386 & 189.6 & 274.7 &  -85.4 &   &   &   &  \\
2016-05-11 & CFHT & 7519.98075 & w &  5 &  600 & 2.426 & 1.423 &  3.511 & 181.3 & 274.7 &  -85.2 & 18.490$\pm$0.002 & 18.901$\pm$0.003 & 19.618$\pm$0.003 & 21.080$\pm$0.006\\
2016-05-12 & CFHT & 7521.00253 & w &  7 &  840 & 2.424 & 1.421 &  3.703 & 173.7 & 274.7 &  -84.9 & 18.502$\pm$0.002 & 18.903$\pm$0.000 & 19.612$\pm$0.002 & 21.079$\pm$0.004\\
2016-05-31 & CFHT & 7539.96859 & w &  5 &  600 & 2.381 & 1.435 & 11.317 & 125.4 & 274.2 &  -79.8 & 18.915$\pm$0.003 & 19.329$\pm$0.003 & 20.048$\pm$0.004 & 21.545$\pm$0.007\\
2016-05-31 & HCT  & 7540.29466 & r$_c$& 5& 625 & 2.381 & 1.435 & 11.317 & 125.4 & 274.2 &  -79.8 & 18.940$\pm$0.015 & 19.342$\pm$0.014 & 20.032$\pm$0.019 & 21.449$\pm$0.064\\
2016-06-05 & CFHT & 7544.95929 & w &  5 &  600 & 2.369 & 1.452 & 13.431 & 122.2 & 274.0 &  -78.4 & 19.095$\pm$0.003 & 19.486$\pm$0.003 & 20.164$\pm$0.004 & 21.627$\pm$0.009\\
2016-06-08 & CFHT & 7547.96720 & w &  7 &  840 & 2.363 & 1.465 & 14.649 & 120.7 & 273.8 &  -77.6 &  &   &   &  \\
2016-06-27 & CFHT & 7566.97056 & w &  1 &  120 & 2.321 & 1.583 & 21.039 & 114.6 & 272.9 &  -72.2 & 19.789$\pm$0.013 & 20.175$\pm$0.015 & 20.858$\pm$0.018 & 22.190$\pm$0.031\\
2016-06-29 & CFHT & 7568.96143 & w & 15 & 1800 & 2.316 & 1.599 & 21.567 & 114.1 & 272.8 &  -71.7 & 19.940$\pm$0.008 & 20.315$\pm$0.009 & 20.990$\pm$0.011 & 22.232$\pm$0.018\\
2016-07-01 & CFHT & 7570.82249 & w & 10 & 1200 & 2.312 & 1.613 & 22.036 & 113.6 & 272.8 &  -71.1 &   &   &   &   \\
2016-07-03 & CFHT & 7572.84000 & w &  4 &  480 & 2.308 & 1.629 & 22.518 & 113.2 & 272.7 &  -70.5 & 19.974$\pm$0.006 & 20.379$\pm$0.007 & 21.072$\pm$0.008 & 22.420$\pm$0.014\\
2016-07-04 & CFHT & 7573.89747 & w &  7 &  840 & 2.305 & 1.638 & 22.760 & 112.9 & 272.7 &  -70.2 & 19.930$\pm$0.004 & 20.338$\pm$0.005 & 21.026$\pm$0.006 & 22.384$\pm$0.010\\
2018-12-12 & CFHT & 8465.139 & w & 5 & 1200 &3.110& 2.515& 16.1&  295.3& 281.8& 168.3& $>25$ \\
2018-12-31 & CFHT & 8484.122 & w & 9 & 2160 &3.118& 2.316& 12.2&  304.1& 281.8& 171.3& $>26$ \\
\hline
\end{tabular}
}%

Notes:
{a: Telescope};
{b: Exposure Mid Julian date-2450000};
{c: Number of exposures};
{d: Total exposure time, [s]};
{e: Heliocentric, geocentric distance [AU], Solar phase angle, [degrees]};
{f: Position of the extended Sun-target radius vector (anti-solar direction) and negative of the heliocentric velocity vector (dust tail orientation) as seen in the plane of the sky measured east from the north celestial pole};
{g: True anomaly, [degrees]};
{h: Limiting mag for non-detection, or total magnitude};
{i: Inferred nucleus radius [km] assuming an albedo of 4\% or 25\%};
{j: Magnitude through aperture diameters of 10000, 7500, 5000, and 2500 km}.
\end{sidewaystable*}